\documentclass[11pt]{article}
\usepackage{moriond,epsfig}
\usepackage{natbib}
\usepackage{journals}
\usepackage[latin1]{inputenc}

\def\be{\begin{equation}}
\def\ee{\end{equation}}
\def\bea{\begin{eqnarray}}
\def\eea{\end{eqnarray}}

\def\etal{{\it et al}.}
\begin{document}
\vspace*{4cm}
\title{$M/L$ in individual galaxies}

\author{Eric Emsellem}

\address{Université de Lyon 1, CRAL, Observatoire
de Lyon, 9 av. Charles André, 69230 Saint-Genis Laval, France ; CNRS,
UMR 5574 ; ENS de Lyon, Lyon, France}

\maketitle\abstracts{In this short report, I briefly review and illustrate different techniques 
   used to derive mass (hence $M/L$) profiles of individual galaxies.
}

\section{Introduction}
The difficulty of determining the mass (of the stars, gas, dark matter) 
in individual galaxies conditions the uncertainty 
on the estimate of the corresponding mass-to-light ratio $M/L$.
Various tracers have been used to probe the radial mass profiles of galaxies 
at scales from tens of kpc down to the inner parsec, 
starting with HI, 
ionised and molecular gas velocity curves (see Sofue \& Rubin~\citep{SR01} and references therein), and including
velocity dispersions and stellar proper motions as revealed in the Near-Infrared. 
High-resolution two-dimensional kinematic maps of the ionised (H$\alpha$) gas in nearby galaxies 
are now routinely obtained \citep{VHHG04, Chemin+06}, and used to better constrain their central 
mass distribution. A battery of new techniques has now been advocated as valid
means for mass determination at large scales, getting help from globular
clusters, planetary nebulae, satellite galaxies, or stellar streams, X ray
distribution, and even lensing systems. In this short review, I illustrate a few
of these techniques in turn.

\begin{figure}
\centerline{\epsfig{figure=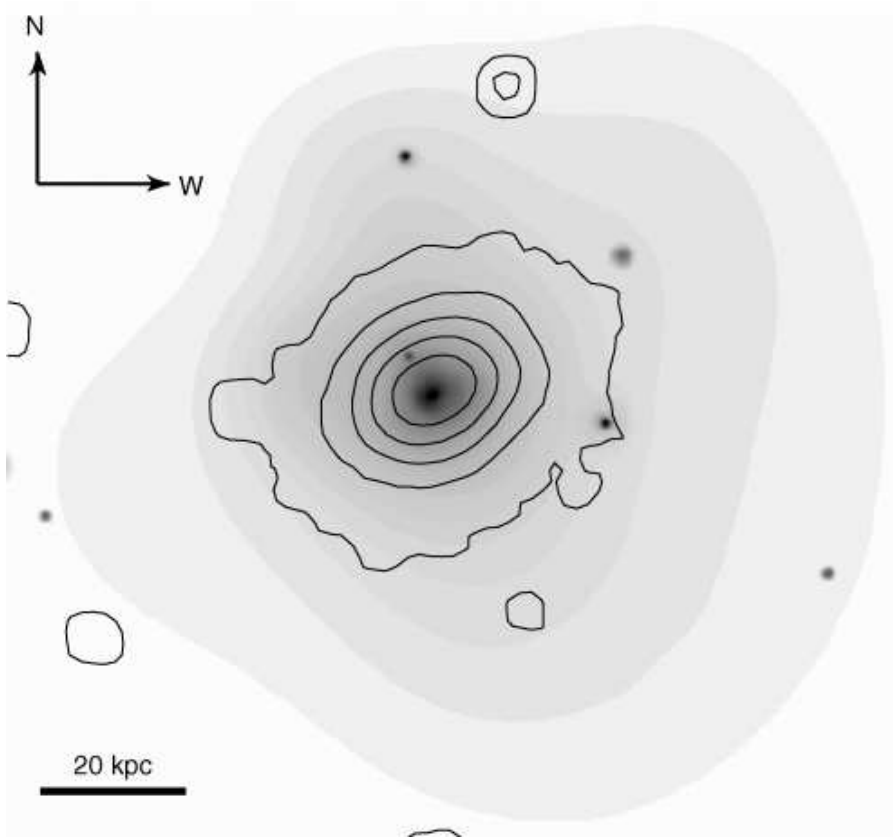,width=12cm}}
\caption{Smoothed X ray image of NGC\,4555, with overlaid optical contours
showing the extent of the stellar component of the galaxy. Extracted from 
O'Sullivan \& Ponman~\cite{OP04}.
\label{fig:Osullivan}}
\end{figure}
\begin{figure}
\centerline{\epsfig{figure=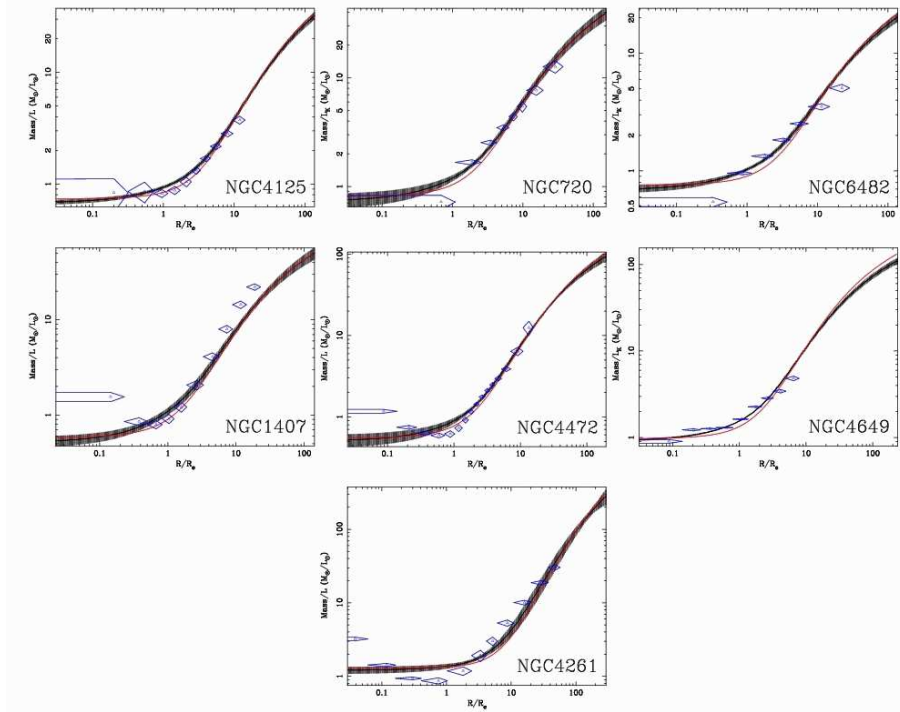,width=12cm}}
\caption{$K$ band mass-to-light ratios for 7 galaxies, obtained by Humphrey
\etal~\cite{Humphrey+06} via Chandra imaging . Extracted from Humphrey \etal~\cite{Humphrey+06}
   (see paper for details).
\label{fig:Humphrey}}
\end{figure}
\section{Masses from X ray halos}
The discovery of X ray halos around bright cluster galaxies led to
its use as a tracer of the gravitational potential typically 
within 5--10 effective radii. Satellites such as 
ASCA and ROSAT allowed global mass measurements which confirmed the dominant
role of the dark matter at these scales, and suggested a relation between the
temperature of the hot gas and the central velocity dispersion of the stellar
component (see \cite{DavisWhite96, LW99}). Such a relation led
Loewenstein \& White~\cite{LW99} to
globally constrain the $M/L$ of a sample of about 30 galaxies in clusters, implying
a relatively low fraction of dark matter within 1~$R_e$ of 20\%, going up to
40--85\% within 6~$R_e$. These measurements are however conditioned by the
assumption of hydrostatic equilibrium, and could be severely affected by
(gravitational) perturbations in the environment of the galaxies. In order to
minimise the effect of the surrounding intra-cluster medium, as well as 
the additional contamination from point-like sources, O'Sullivan \& Ponman~\cite{OP04} studied the
isolated elliptical galaxy NGC~4555 with the ACIS/Chandra instrument (Fig.~\ref{fig:Osullivan}), and found
a large $M/L_B$ of about 57 at 50~kpc implying a very significant fraction of
dark matter.

Detailed mass profiles were subsequently derived by Fukuzawa \etal~\cite{Fukazawa+06} for a
sample of 53 early-type galaxies via data retrieved from the Chandra archive.
They first found relatively good agreement with mass profiles within 1~$R_e$
derived either from stellar population studies or dynamical modelling.
Fukuzawa \etal~\cite{Fukazawa+06} then emphasised the apparent dichotomy in the temperature
profiles between X-ray luminous and dim galaxies, with the former exhibiting a
temperature increasing with radius, while the latter show declining or flat profiles.
The derived mass-to-light ratios seemed to be constant within the central 1~$R_e$, but clearly
rise up outwards (but see e.g., Pellegrini~\etal~\cite{Pellegrini+06}). Similar results were obtained by Humphrey~\etal~\cite{Humphrey+06} for 7
nearby ellipticals, with the central $M/L_K$ being consistent within 1~$R_e$
with a Kroupa initial mass function (IMF; Fig.~\ref{fig:Humphrey}). 

\section{Lensing}
\begin{figure}
\centerline{\epsfig{figure=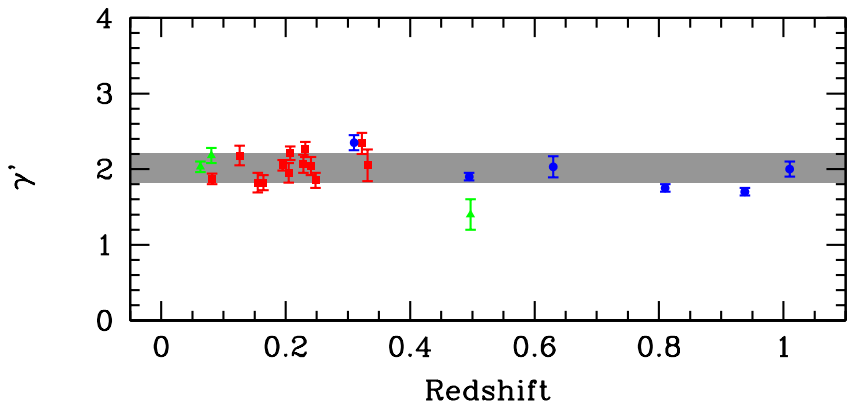,width=12cm}}
\caption{Logarithmic density slope $\gamma$ of field galaxies plotted against
redshift. The grey box indicates the rms spread of 0.19. Extracted from
Koopmans \etal~\cite{Koopmans+06} (see paper for details).
\label{fig:Koopmans}}
\end{figure}
As mentioned above, the main uncertainties in the determination of mass profiles
from X ray studies result from the assumption of a spherical halo in
hydrostatic equilibrium as well as from the difficulty to remove the background
and point-like sources. Any technique has its own limitations, and it is
therefore important to test various methods to gain confidence in the obtained results.
Strong lensing is a gravitational effect and is therefore a natural path to
probe mass in galaxies. This requires first accurate and high signal-to-noise
imaging of the system, and usually assumes an a priori form for the underlying lens 
(e.g., isothermal sphere with $\rho \propto r^{-2}$). Such a technique has
recently been mastered by Rusin \& Kochanek~\cite{RK05} who used a sample of 22 lenses and the
constraints provided by the fundamental plane to probe the mass profiles of
early-type galaxies. As this study applies to the global sample of galaxies, it
additionally assumes homology and a similar history. Optimising for the
logarithmic density slope $\gamma$, Rusin \& Kochanek~\cite{RK05} find that their sample of galaxies is
better represented by nearly isothermal profiles, with an evolution of $M/L_B$
with redshift given by $d \log{(M/L_B)} / dz = -0.5 \pm 0.19$, consistent with,
     e.g., the
previous constraint obtained by van~Dokkum~\etal~\cite{vDFKI98} of $-0.8 < d \log{(M/L_B)} / dz < -0.4$.
A Salpeter IMF would then requires a mean star formation redshift at $z > 1.5$.

A similar analysis but more detailed analysis was performed by Treu~\etal~\cite{Treu+06} and Koopmans~\etal~\cite{Koopmans+06} who 
made a joint stellar dynamical and strong lensing analysis of 15 early-type galaxies with redshift
$0.06 < z < 0.33$ (Sloan Lens ACS Survey, SLACS). Velocity dispersions $\sigma$ were obtained from the SDSS
project, and ACS/HST data were used to derive the lensing parameters. After
deriving the mass within the Einstein radius, the Jeans Equations were solved assuming isotropy
and a density profile of the form $\rho \propto r^{- \gamma}$, to compare the expected
dispersion with the measured SDSS values. They finally solved for the combined
probability to estimate $\gamma$, and found slopes still consistent with isothermal
spheres ($\gamma = 2$) within $R_e / 2 $, with no significant evolution with
redshift (Fig.~\ref{fig:Koopmans}).

Weak lensing studies were also used by e.g., Hoekstra~\etal~\cite{Hoekstra+05}, to constrain the Virial
mass $M_{vir}$ of galaxies. The measured signal probe then the average properties for a
sample of relatively isolated galaxies, still allowing to examine the behaviour
of mass in 7 bins of luminosity. Their results are consistent with a scaling of 
$M_{vir}\propto L^{1.5}$, and a lower stellar mass fraction in earlier-type
galaxies.

\section{Velocity profiles}
One of the most standard technique for the determination of mass profiles relies
on the abundant HI content of disk galaxies and its observed kinematics (see \citep{SR01}
and references therein). Some concerns were raised by the lack of spatial
resolution in the central region which impairs the application of decomposition
techniques to constrain the relative contribution of the stars, gas and dark
matter. The use of H$\alpha$ mapping (via Fabry-Perot interferometers) 
seems to properly address this issue, although the
complex dynamics expected in the central regions of spiral (and barred) galaxies
enters then as an extra complication. A remarkable sample of 329 H$\alpha$ rotation curves 
of field spirals was obtained by Vogt~\etal~\cite{VHGH04}. I should also mention the unique
coverage of Virgo spirals conducted by Chemin~\etal~\cite{Chemin+06}, which will permit a
detailed study of the mass profiles as well as the impact of environment on the
gas content, distribution and kinematics.

Two-dimensional maps are certainly a requirement if we wish to disentangle
the global (circular?) motion from the effect of density waves. Via the combined 
use of high resolution ionised (H$\alpha$) and molecular (CO) gas velocity
fields, Simon~\etal~\cite{SBLB03} constrained the mass variation with radius in the
dwarf spiral NGC\,2976. The addition of multi-colour optical and near-infrared
images allowed the authors to suggest that the stellar mass fraction is relatively high in the
central region, with the dark matter having then a profile shallower than $\rho
\propto r^{-0.17}$, with the caveat that the obtained central $M/L_K$ may be 
too low to be accounted by normal stellar populations.
A similar study, but this time of a flocculent isolated spiral NGC\,4414, had
been conducted by Vallejo~\etal~\cite{VBB03}, who combined high resolution CO data
with extended HI rotation curve, to conclude that the mass-to-light ratio cannot
be constant throughout the galaxy, but that dark matter has certainly a nearly
negligible role in the central 7~kpc.

\begin{figure}
\centerline{\epsfig{figure=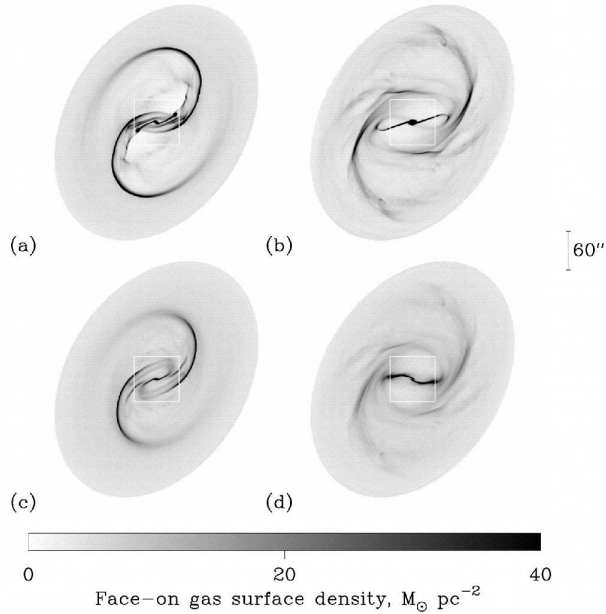,width=8cm}}
\caption{Gas response for different models of NGC\,4123: these models are used
to constrain the pattern speed of the bar as well as the mass-to-light ratio.
Extracted from Weiner~\etal~(see \cite{WSW01} and \cite{WWvGS01} for details).
\label{fig:Weiner}}
\end{figure}
\section{Bars, warps and rings}
Density waves can in fact be a rather efficient tool to constrain the mass
distribution in galaxies. By examining the gas response to the potential of a
tumbling bar, Weiner~\etal~\cite{WSW01,WWvGS01} studied the pattern speed $\Omega_p$ and $M/L$
of the bar in NGC\,4123 (Fig.~\ref{fig:Weiner}). The presence of strong shocks is clearly revealed in their
Fabry-Perot H$\alpha$ maps and this significantly narrows down the range of allowed 
values for $\Omega_p$ and $M/L$. The study of the warp in NGC\,5055 Battaglia~\etal~\cite{BFOS06} was
not as successful, as it does not seem to bring any strong constraint on the
presence of dark matter in the outer part of that galaxy.
This contrasts with the use of outer HI rings in early-type galaxies, as
emphasised by Franx, van Gorkom \& de Zeeuw~\cite{FvGdZ94} who showed that the gravitational potential is
nearly circular in the plane of the observed ring, and that there is a
significant increase of $M/L$ with radius. A new case, namely the early-type disc galaxy NGC\,2974,
is now under study by Weijmans~\etal\ using both HI data for
the outer ring and integral-field spectroscopic {\tt SAURON} data for the inner
ionised gaseous component. 

\section{Stellar dynamics}
Stellar kinematics has also been extensively used with the help of sophisticated
dynamical models to constrain the mass-to-light ratios of nearby galaxies 
up to a few effective radii. The apparent simplicity implied by the extension
and smoothness of the stellar component is unfortunately accompanied by the
potential richness of the corresponding orbital structure. Strong assumptions 
on the geometry and dynamics of the systems are therefore, and once again, 
required to progress on this front. Even for simple spherical systems, there is
a long known degeneracy between the anisotropy and the mass profile \cite{BM82}.
Sanchis~\etal~\cite{SLM04} have thus recently emphasised the importance of higher order velocity moments
(and more specifically of the even moments) to break this degeneracy. Kronawitter~\etal~\cite{KSGB00}
included the velocity dispersion but also the fourth Gauss-Hermite term $h_4$ in their study of a small
sample of early-type galaxies. Spherical models based on distribution function
components showed that the $M/L$ is increasing outwards in these galaxies
with however rather standard values in the inner parts (consistent with observed
stellar populations). They also convincingly showed that the derived increase in $M/L$
is consistent with the one inferred from X ray halo studies.

\begin{figure}
\centerline{\epsfig{figure=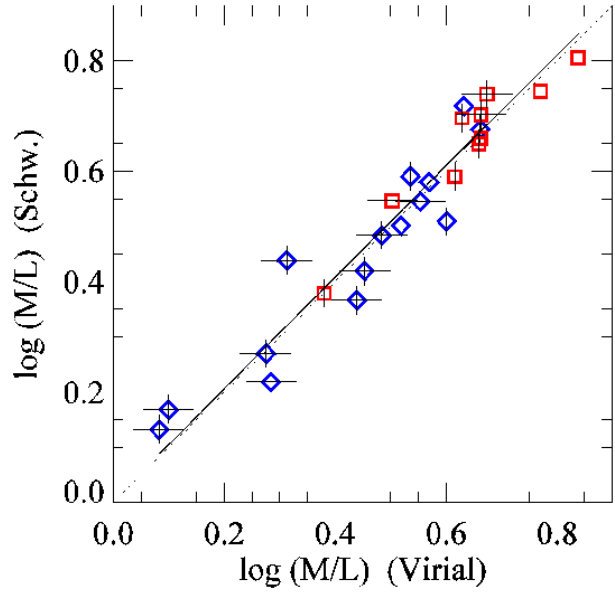,width=8cm}}
\caption{Dynamical $M/L$ (as derived from Schwarzschild models) versus Virial $M/L$
(derived from stellar kinematics) for a sample of 25 early-type galaxies obtained by 
Cappellari~\etal~\cite{Cappellari+06}. Extracted from \cite{Cappellari+06} (see
paper for details).
\label{fig:Cappellari}}
\end{figure}
\begin{figure}
\centerline{\epsfig{figure=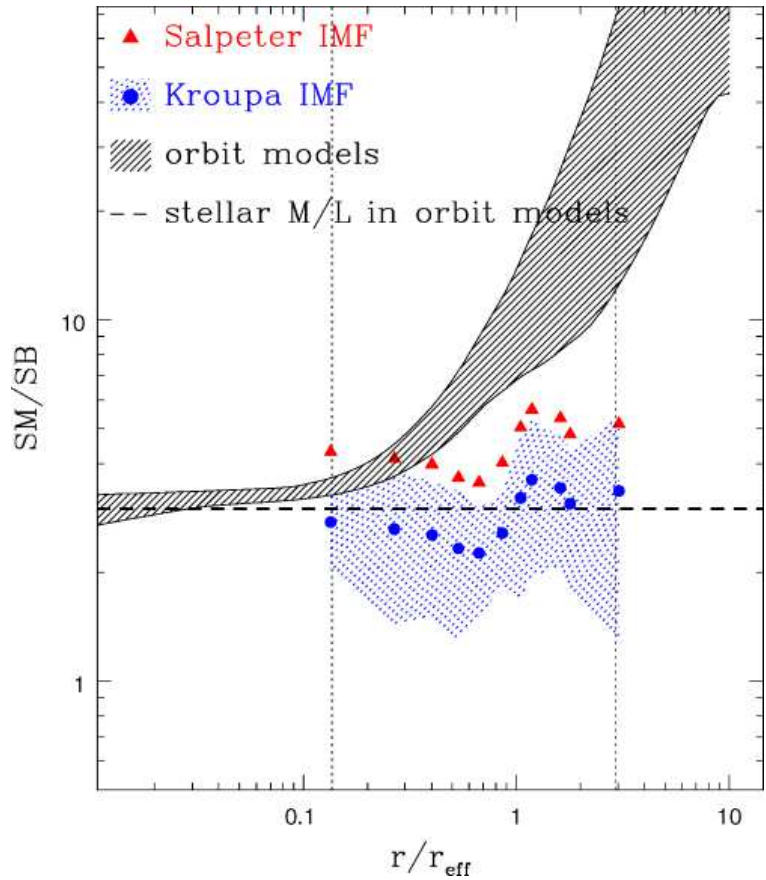,width=10cm}}
\caption{Surface mass versus brightness for the Coma galaxy NGC\,4807 as
predicted by the Schwarzschild model (grey area) and compared to the one
expected from a Salpeter (red line) or Kroupa (blue line) IMF. Extracted from Thomas~\etal~\cite{ThomasJ+05}
(see paper for details).
\label{fig:JThomas}}
\end{figure}

The first significant attempt at deriving stellar $M/L$ in flattened early-type galaxies
was achieved by van der Marel~\cite{vanderMarel91}, who solved the Jeans equations to model a
sample of elliptical and lenticular galaxies. A clear trend of $M/L$
increasing with luminosity was obtained, although with a relatively large scatter.
The advent of more general modelling techniques, such as the Schwarzschild
method \cite{Schwarzschild79,Schwarzschild82,RT88} and its application to high
quality stellar kinematics \cite{vdMCdZR98,Rix+97,Verolme+02} led to impressive progresses
in our understanding of the dynamical structure of nearby galaxies 
\cite{Gebhardt+03,VME04,ThomasJ+05}. One nagging issue was the
recurrent appearance of the degeneracy between black hole mass and $M/L$, 
clearly illustrated in the first ambitious study of this kind by Gebhardt~\etal~\cite{Gebhardt+03}, 
which thus require both high spatial resolution and large-scale spectroscopic data.
The even greater need for two-dimensional spectroscopy to obtain realistic
models of galaxies was subsequently emphasised \cite{CE04,Krajnovic+05}. An
apparent degeneracy of the models with respect to the often unknown inclination parameter was
also suggested by Krajnovi{\'c}~\etal~\cite{Krajnovic+05}, the mass-to-light ratio not being
significantly affected by this issue.

Following this path, Thomas~\etal~\cite{ThomasJ+05} obtained a rather strong constraint on the
presence of dark matter at a few effective radii of the Coma galaxy NGC\,4807 with an increase in $M/L$ of
more than an order of magnitude between 1 and 5 $R_e$ (Fig.~\ref{fig:JThomas}).
The stars are, however, still dominant within 1~$R_e$ as confirmed by Cappellari~\etal~\cite{Cappellari+06}
using state-of-the-art Schwarzschild models and integral-field data. This work
also showed that the maximum contribution from non-homology of early-type
galaxies to the tilt of the fundamental plane is about 6\%, as the dynamical $M/L$ 
(from dynamical models) and the $M/L$ derived from the Virial theorem (as predicted using the
stellar kinematic maps) agree amazingly well with each other (Fig.~\ref{fig:Cappellari}). This suggests that
the tilt of the FP is dominated by a true variation of the $M/L$ from galaxy to
galaxy. A comparison of the dynamical $M/L$ and the stellar $M/L$ (obtained via
stellar line indices) finally suggests that massive slowly rotating galaxies have a
larger fraction of dark matter than fainter ones, with an upper limit of about 30\%
within 1~$R_e$ \cite{Cappellari+06}.

\section{The vanishing}
Romanowsky~\etal~\cite{Romanowsky+03} recently probed the outer regions of
NGC\,3379 up to 3~$R_e$ using the Planetary Nebulae Spectrograph finally concluding 
that dark matter was not required to reproduce the radial dispersion measurements.
This result was severely questioned by Dekel~\etal~\cite{Dekel+05}, who reminded
us of the density/anisotropy degeneracy. They showed that a bias towards
radial stellar orbits may be naturally expected at a few effective radii, which
would produce a dispersion profile similar to the one observed in NGC\,3379
even though the mass profile is dominated by dark matter.
A recent, but still preliminary, measurement of the stellar velocity dispersion at $\sim 3 R_e$
using the {\tt SAURON} spectrograph by Weijmans~\etal\ indicates that the bulk stellar
population has in fact a rather constant dispersion profile, which would, if
confirmed, favour the presence of a significant amount of dark matter, and
reveal a significant discrepancy between different tracers.

\section{Satellites, X-ray binaries and stellar populations}
By studying the dynamics of small satellites around larger galaxies it is in
fact possible to restrict the allowed parameter space for a set of mass models.
This was achieved for the Andromeda galaxy by Geehan~\etal~\cite{GFBG06}
who analysed the stellar streams around M\,31 and 
found a rather normal $M/L$ for its disc and bulge.
The original approach of Dehnen \& King~\cite{DK06} included the
study of the motions of 5 X-ray binaries (LMXB) detected in the Sculptor dwarf spheroidal. 
The rare occurrence of such binaries suggests that Sculptor must have retained all its LMXBs.
Taking into account a visible mass of $10^7$~M$_{\odot}$, and a measured
stellar velocity dispersion of only 11~km.s$^{-1}$, the mass
distribution of Sculptor should be dominated by a dark halo of 10$^{9}$~M$_{\odot}$
within 1.5~kpc, which would imply a total $M/L$ of at least a few hundreds.
This tentative result is one of the very few constraints we have on the mass
distribution of dwarfs, but we should realise that it will be a challenge to
confirm the assumptions on which it is based.
 
I should add the mention of two recent illustrative works where
stellar mass-to-light ratios of nearby dwarfs were derived via spectral line
indices \cite{GGvdM05,Thomas+06}. Central counter-rotating core were revealed in
both NGC\,770~\cite{GGvdM05} and VCC\,510~\cite{Thomas+06}, their spectral
signature and stellar synthesis models indicating metal poor nuclei,
the possible remnants of a merger/accretion event.

I will finally end up with a brief word on Modified Newton Dynamics (MOND).
Apart from the fruitful debate it triggered\footnote{A personal opinion,
obviously}, MOND is at least useful for one purpose: as emphasised by McGaugh~\cite{McGaugh05}, 
it proposes that we should view the mass discrepancy in the outer parts of galaxies as
an empirical relation. This empirical relation is a very useful "tracer", which 
can then be analysed in the light of stellar population models, dark matter, MOND or any preferred
prescription.

\section{Brief conclusions}
Baryons seem to dominate the central parts of galaxies (within 1--2~$R_e$),
probably even in dwarfs, and stellar mass-to-light ratios are consistent with 
the dynamical estimates in these regions. The mass discrepancy increases
outwards, which suggests the presence of dark matter, including in
early-type galaxies. The outer mass profiles are roughly consistent with NFW
profiles, but these seem to fail to reproduce the central, shallow, density slopes.

Each technique mentioned above represents one path to the determination of mass profiles
in individual galaxies. They all have their own specific (nagging) issues, and include many 
geometrical, physical, numerical assumptions which should be carefully examined
in turn. As emphasised, point-like sources, background flux, counts and the
assumption of hydrostatic equilibrium have to be considered when using X-ray spectroscopy and imaging. 
Strongly lensed systems are
rare, and biased toward "relatively" high redshifts and large masses. HI,
H$\alpha$, and CO are clearly complementary, but we always need to examine 
large-scale perturbations, as well as the assumptions we make on the gas dynamics.
Outer gas rings are efficient tools in this context but only probe the potential locally, and are,
unfortunately, very rare. Density waves such as bars, spirals and warps are
good tracers of the dynamics of the system (hence of its mass), but may not bring
very strong constraints, and are usually restricted to the central regions.
Stellar dynamics is complex and require state-of-the-art modelling with well
tuned assumptions. Still, degeneracies exist which may sometime blur the robustness of
the results. Tracers such as planetary nebulae, globular clusters or streams are
obvious targets, but it is not clear if these are effectively good tracers and
if basic assumptions such as stationarity can be applied. Finally $M/L$ derived
from stellar populations is probably the least robust parameter provided by
spectral synthesis models, although we witness steady progress there. 
More generally, I think we can be relatively optimistic as
we today follow the advent of new instruments, techniques, and numerical codes which
clearly address the issues mentioned here. We then need to find out if such 
vastly different approaches deliver consistent answers when probing the same mass regimes. 

\section*{Acknowledgements}
I would like to thank the organisers for providing me with 
the opportunity to participate to such a fruitful conference.
I would also like to warmly thanks colleagues who helped me prepare this review,
namely Michele Cappellari, Roelof de Jong, Harald Kuntschner,
Stacy McGaugh, Gary Mamon, Claudia Maraston, Tom Statler and Anne-Marie Weijmans.

\bibliography{Emsellem_Moriond06}

\end{document}